\documentclass[runningheads]{svmult}

\usepackage{makeidx}   
\usepackage{graphicx}  
\usepackage{subeqnar}  
\usepackage{multicol}  
\usepackage{physprbb}  
\makeindex             

\def\etal{{\em et~al.\ }}
\def\spose#1{\hbox to 0pt{#1\hss}}
\def\lta{\mathrel{\spose{\lower 3pt\hbox{$\mathchar"218$}}
     \raise 2.0pt\hbox{$\mathchar"13C$}}}
\def\gta{\mathrel{\spose{\lower 3pt\hbox{$\mathchar"218$}}
     \raise 2.0pt\hbox{$\mathchar"13E$}}}
\def\kms{{\rm\,km\,s^{-1}}}

\def\kpc{{\rm\,kpc}}
\def\mpc{{\rm\,Mpc}}

\begin{document}
\title*{Dynamical Masses of Elliptical Galaxies
       }
\toctitle{Dynamical Masses of Elliptical Galaxies}

%
%
\titlerunning{Dynamical Masses of Elliptical Galaxies}
%
\author{Ortwin Gerhard
}
\authorrunning{Ortwin Gerhard}
%
%
\institute{Astronomisches Institut, Universit\"at Basel,
Venusstrasse 7, 4102 Binningen, Switzerland
}

\maketitle              

\begin{abstract}
  Recent progress in the dynamical analysis of elliptical galaxy
  kinematics is reviewed. Results reported briefly include (i) the
  surprisingly uniform anisotropy structure of luminous ellipticals,
  (ii) their nearly flat (to $\sim 2R_e$) circular velocity curves,
  (iii) the Tully-Fisher and $M/L - L$ relations and the connection to
  the Fundamental Plane, and (iv) the large halo mass densities
  implied by the dynamical models.
\end{abstract}

\section{Introduction}
Elliptical galaxies are the most massive galaxies and are generally
found in dense environments. In the context of hierarchical models,
they represent an advanced step in the galaxy formation process. Their
mass distributions and dark halo properties, while not yet as
well-understood observationally as those of spiral galaxies, are of
considerable interest. X-ray and gravitational lensing studies
imply mass-to-light ratios $M/L\sim 100$ on scales of $\sim 100
\kpc$ for the most massive ellipticals containing hot gas atmospheres.
Here the new XMM and Chandra satellite data will lead to mass
measurements of unprecedented resolution and accuracy.

In the central $\sim 2R_e$, mass distributions $M(r)$ may be
determined from absorption-line profile measurements and dynamical
models.  Planetary nebulae and globular clusters can be used as
discrete velocity tracers to $\sim 4-5R_e$ (typical effective radii
$R_e$ are in the range 3-10 kpc). In less massive, gas-poor systems
these may give the main constraints on $M(r)$. In gas-rich ellipticals
where $M(r)$ can be accurately determined from the X-ray gas emission,
the stellar-kinematic data will constrain the orbit structure best.

The following sections give a brief review of the dynamical analysis
of the stellar-kinematic data, and the results obtained so far on mass
distributions in ellipticals and on their dynamical family properties.

\section{Dynamical mass estimation}

As is well known, velocity dispersion profile measurements (and
streaming velocities, if the galaxy rotates) do not suffice to
determine the distribution of mass with radius, due to the degeneracy
with orbital anisotropy. Only with very extended measurements can a
constant $M/L$ model be ruled out (e.g., Saglia \etal 1993), but even
then the detailed $M(r)$ remains undetermined. Absorption line profile
shapes (giving line-of-sight velocity distributions $L(v)$, LOSVD for
short), however, contain additional information with which this
degeneracy can largely be broken. Simple spherical models are useful
to illustrate this (Gerhard 1993): at large radii, radial orbits are
seen side-on, resulting in a peaked LOSVD (positive Gauss-Hermite
parameter $h_4$), while tangential orbits lead to a flat-topped or
double-humped LOSVD ($h_4<0$). Similar considerations can be made for
edge-on or face-on disks (Bender 1990, Magorrian \& Ballantyne 2001)
and spheroidal systems (Dehnen \& Gerhard 1993).

One may think of the LOSVDs constraining the anisotropy, after which
the Jeans equations can be used to determine the mass distribution.
However, the gravitational potential influences not only the widths,
but also the shapes of the LOSVDs (see illustrations in Gerhard 1993).
Furthermore, eccentric orbits visit a range of galactic radii and may
therefore broaden a LOSVD near their pericentres as well as leading to
outer peaked profiles. Thus, in practise, the dynamical modelling to
determine the orbital anisotropy and gravitational potential must be
done globally, and is typically done in the following steps:

(0) choose geometry (spherical, axisymmetric, triaxial);

(1) choose dark halo model parameters, and set total luminous plus
    dark matter potential $\Phi$;

(2) write down a composite distribution function (DF)
    $f=\Sigma_k a_k f_k$, where the $f_k$ can be orbits, or
    DF components such as $f_k(E,L^2)$, with free $a_k$; 

(3) project the $f_k$ to observed space, $p_{jk}=\int K_j f_k d\tau$,
    where $K_j$ is the projection operator for observable $P_j$, and
    $\tau$ denotes the line-of-sight coordinate and the velocities; 
    
(4) fit the data $P_j = \Sigma_k a_k p_{jk}$ for all observables
    $P_j$ simultaneously, minimizing a $\chi^2$ or negative
    likelihood, and including regularization to avoid spurious large
    fluctuations in the solution. This determines the $a_k$, i.e, the
    best DF $f$, given $\Phi$, which must be $f>0$ everywhere;

(5) vary $\Phi$, go back to (1), and determine confidence limits
    on the parameters of $\Phi$. \\
Such a scheme was employed, e.g., using orbits in spherical potentials by
    Rix \etal (1997) and Romanowsky \& Kochanek (2001); using DF
    components in spherical potentials by Gerhard \etal (1998), Saglia
    \etal (2000), and Kronawitter \etal (2000); with orbits in
    axisymmetric geometry by Cretton \etal (2000) and Gebhardt \etal
    (2000); and with DF components in axisymmetry by
    Matthias \& Gerhard (1999) and Statler \etal (1999).  The
    modelling techniques used to constrain black hole masses from
    nuclear kinematics and dark halo parameters from extended
    kinematics are very similar.

Line-profile shape parameter measurements are now available for many
nearby ellipticals (e.g., Bender, Saglia \& Gerhard 1994), but those
reaching to $\sim 2 R_e$ are still scarce (see Kronawitter \etal
2000).  Modelling of the mass profiles of ellipticals from such data
has been done for some two dozen round galaxies in the spherical
approximation, and for a few cases using axisymmetric three-integral
models.

Typically the models that fit individual galaxies best imply small to
modest amounts of dark matter within $2R_e$, but there are ellipticals
which are very well represented by constant $M/L$ models out to these
radii. Most results obtained so far are from spherical models for round
ellipticals; their mass distributions and radially anisotropic orbit
structure are discussed in the next section.  The effects of intrinsic
deviations from sphericity and of embedded, near-face-on disks have
been discussed by Kronawitter \etal (2000), Magorrian \& Ballantyne
(2001), and Gerhard \etal (2001).  Axisymmetric models exist only for
very few ellipticals. In NGC 1600 (E3.5, Matthias \& Gerhard 1999),
NGC 2300 (E2, Kaeppeli 1999), NGC 2320 (E3.5, Cretton, Rix \& de Zeeuw
2000), and NGC 3379 (E1, Gebhardt \etal 2000) radially
anisotropic structure at $\sim 0.5 R_e$ has been inferred from
three-integral models along the major axis, similar to that found
in the round galaxies, but with less anisotropy on
the minor axis.

Because the dark matter fraction inside $2R_e$ is still modest, and
the orbit structure in the outer main bodies of ellipticals is not
well-constrained by data that end at $2R_e$, it will be important to
include discrete velocity data from planetary nebulae (PN) or globular
clusters (GC) that reach to larger radii. Such data can be included
directly in the above modelling scheme, using a likelihood
maximization (Romanowsky \& Kochanek 2001), or can be used a
posteriori to differentiate between dynamical models at radii beyond
the absorption line data on which these are based (Saglia \etal 2000).
The number of ellipticals with such data is still small, but is
expected to be growing rapidly thanks to the special purpose PN
spectrograph (Douglas \etal 2002) and ongoing programs with
globular cluster samples around ellipticals. Typically one may expect
a few hundred discrete velocities per galaxy from these programs.
Using slitless spectroscopy at the VLT M\'endez \etal (2001) succeeded
in measuring 535 PN radial velocities around the E5 galaxy NGC 4697.
Detailed modelling of these data is still in progress, but simple
models for this relatively low luminosity elliptical galaxy are
consistent with constant $M/L$ out to $\sim 3R_e$, in agreement with
Dejonghe \etal (1996).

Two particularly interesting cases are the central galaxies of the
Fornax and Virgo clusters, NGC 1399 and M87. From a comparison of
dynamical models for the absorption line kinematics (Saglia \etal
2000) with the mass distribution obtained from ASCA data (Ikebe \etal
1996), it appears that PN (Arnaboldi \etal 1994) and GC velocities
(Richtler \etal 2001) are just in the right radial range to allow a
study of the transition between the potential of the central NGC 1399
galaxy and the potential of the Fornax cluster. Similarly, from a
study of the stellar kinematics and the GC velocities around M87, and
a comparison to the X-ray mass profile, Romanoswky \& Kochanek (2001)
find a rising circular velocity curve, and suggest that the potential
of the Virgo cluster may already dominate at $r \sim 20 \kpc$ from
the center of M87.

\section{Dynamical family properties} 

The discussion in this section is based on the work of Kronawitter
\etal (2000) and Gerhard \etal (2001), who analyzed the line-profile
shapes of a sample of 21 mostly luminous, slowly rotating, and nearly
round elliptical galaxies in a uniform way, using spherical dynamical
models. A similar study using three-integral axisymmetric models will
be very worthwhile, but is still some time away. The sample of
Kronawitter \etal includes a subsample with mostly new extended
kinematic data, reaching to $\sim 2R_e$, and a subsample based on the
less extended older data of Bender \etal (1994). Based on these data
and on photometry, non-parametric spherical models were
constructed from which circular velocity curves, anisotropy profiles,
and radial profiles of $M/L$ were derived, including confidence
ranges. The main results from this study are as follows.

(1) The circular velocity curves (CVCs) of elliptical galaxies are
flat to within $\simeq 10\%$ for $R\gta 0.2R_e$ to at least $R\gta
2R_e$, independent of luminosity (Fig.~1).  The CVC is a convenient
measure of the potential even though luminous elliptical galaxies do
not rotate rapidly. Quantitatively, the median ratio of the circular
velocity at the radius of the outermost kinematic data point,
$v_c(R_{\rm max})$, to the maximum circular velocity of the respective
best model, $v_c^{\rm max}$, is 0.94, with $95\%$ confidence ranges of
$\sim \pm 0.1$. This argues against strong luminosity segregation in
the dark halo potential.

(2) Despite the uniformly flat CVCs, there is a spread in the ratio of
the CVCs from luminous and dark matter, i.e., in the radial variations
of cumulative mass-to-light ratio. The sample includes galaxies with
no indication for dark matter within $2R_e$, and others where the best
dynamical models result in local $M/L_B$s of 20-30 at $2R_e$.  As in
spiral galaxies, the combined rotation curve of the luminous and dark
matter is flatter than those for the individual components
(``conspiracy'').

(3) Most of these ellipticals are moderately radially anisotropic,
with average $\beta\equiv 1-\sigma_\theta^2/\sigma_r^2\simeq 0-0.35$,
again independent of luminosity. The dynamical structure of
ellipticals is therefore surprisingly uniform.  The maximum circular
velocity is accurately predicted by a suitably defined central
velocity dispersion. Averaging $\sigma$ within $0.1R_e$,
$\sigma_{0.1}=0.66 v_c^{\rm max}$.

(4) Elliptical galaxies follow a Tully-Fisher (TF) relation with
marginally shallower slope than spiral galaxies (see also Magorrian \&
Ballantyne 2001). At given circular velocity, they are about 1 mag
fainter in B and about 0.6 mag in R, and appear to have slightly lower
baryonic mass than spirals, even for the maximum $M/L_B$ allowed by
their kinematics (minimum dark halo models).  The residuals from the
TF (and Fundamental Plane, FP) relations do not correlate with
dynamical anisotropy $\beta$.

\begin{figure}[t]
\begin{center}
\includegraphics[width=.8\textwidth]{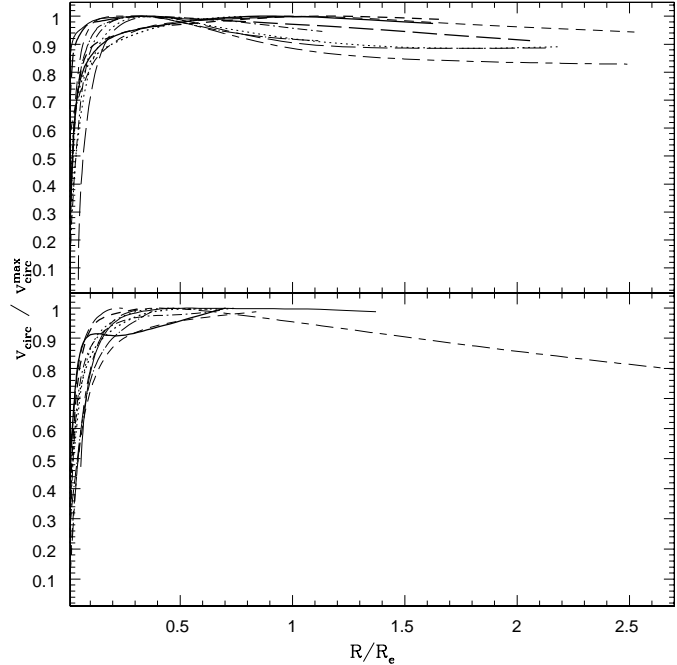}
\end{center}
\caption[]{``Best model'' circular velocity curves of all galaxies
from the Kronawitter \etal (2000) sample, plotted as a function of
radius scaled by the effective radius $R_e$, and normalized by
the maximum circular velocity. The upper panel shows the galaxies
from the extended kinematics subsample, the lower panel the galaxies
from the subsample with older data from Bender \etal (1994). From
Gerhard \etal (2001).}
\label{eps1}
\end{figure}

(5) The luminosity dependence of $M/L_B$ indicated by the tilt of the FP
corresponds to a real dependence of dynamical $M/L_B$ on $L_B$. The tilt
of the FP is therefore not due to deviations from homology or a
variation of dynamical anisotropy with $L_B$, although the slope of
$M/L_B$ versus $L_B$ could still be influenced by photometric
non-homology.  The tilt can also not be due to an increasing dark matter
fraction with $L_B$, unless (i) the most luminous ellipticals have a
factor $>3$ less baryonic mass than spiral galaxies of the same
circular velocity, and (ii) the range of IMF is larger than currently
discussed, and (iii) the IMF or some other population parameter
varies systematically along the luminosity sequence such as to undo
the increase of $M/L_B$ expected from simple stellar population models
for more metal-rich luminous galaxies. This seems highly unlikely.

(6) The tilt of the FP is therefore best explained as a stellar
population effect.  Population models show that the values and the
change with $L_B$ of the maximal dynamical $M/L_B$s are consistent
with the stellar population $M/L_B$s based on published metallicities
and ages, within the uncertainties of IMF and distance scale.  Because
of (5) above and because the observed correlation between age and
luminosity is far too weak to explain the $M/L_B-L_B$ relation (Forbes
\& Ponman 1999), the main driver of the B-band tilt is therefore
probably metallicity.  This would not explain the observed K-band
tilt, which must then be explained by a secondary population effect.

(7) The population models show that the dynamical models would have
overestimated the luminous masses of these elliptical galaxies by as
much as a factor $\approx 2$ only if (i) the flattest IMFs at low
stellar masses discussed for the Milky Way are applicable, and
simultaneously (ii) a short distance scale ($H_0\simeq
80\kms\mpc^{-1}$) turns out to be correct.  For lower values of $H_0$
and/or other IMFs the difference is smaller. Together with (4) this
makes it likely that elliptical galaxies have indeed nearly maximal
$M/L_B$ ratios (minimal halos).

(8) In the models with maximum stellar mass, the dark matter contributes
$\sim 10-40\%$ of the mass within $R_e$. The flat CVC
models, when extrapolated beyond the range of kinematic data, predict
equal interior mass of dark and luminous matter at $\sim 2-4R_e$,
consistent with results from X-ray analyses.

(9) Even in maximum stellar mass models, the halo core densities and
phase-space densities are at least $\sim 25$ times larger and the halo
core radii $\sim 4$ times smaller than in maximum disk spiral galaxies
with the same circular velocity (Persic, Salucci \& Stel 1996).
Correspondingly, the increase in $M/L$ sets in at $\sim 10$ times
larger acceleration than in spirals.  This could imply that elliptical
galaxy halos collapsed at high redshifts or perhaps even that some of
the dark matter in ellipticals might be baryonic.

%

\end{document}